\documentclass[prd,twocolumn,showpacs,amsmath,amssymb,floatfix,nofootinbib]{revtex4}
\usepackage{graphicx,color,dcolumn,booktabs}
\usepackage{longtable,lscape}

\usepackage{indentfirst}

\def\lrpartial{\buildrel\leftrightarrow\over\partial}

\begin{document}
\title{Understanding the branching ratios of $\chi_{c1}\to\phi\phi,\,\omega\omega,\,\omega\phi$ observed at BES-III}

\author{Dian-Yong Chen$^{1,3}$}
\author{Jun He$^{1,3}$}
\author{Xue-Qian Li$^4$}
\author{Xiang Liu$^{1,2}$\footnote{Corresponding author}}\email{xiangliu@lzu.edu.cn}
\affiliation{$^1$Research Center for Hadron and CSR Physics,
Lanzhou University $\&$ Institute of Modern Physics of CAS, Lanzhou 730000, China\\$^2$School of Physical Science and Technology, Lanzhou University, Lanzhou 730000,  China\\
$^3$Institute of Modern Physics, Chinese Academy of Sciences, Lanzhou 730000, China\\
$^4$Physics department, Nankai University, Tianjin 300071, China}

\date{\today}
\begin{abstract}

In this work, we discuss the contribution of the mesonic loops to the
decay rates of $\chi_{c1}\to \phi\phi,\,\omega\omega$ which are
suppressed by the helicity selection rules and $\chi_{c1}\to
\phi\omega$ which is a double-OZI forbidden process. We find that
the mesonic loop effects naturally explain the clear signals of $\chi_{c1}\to \phi\phi,\,\omega\omega$ decay modes observed by the BES collaboration.
Moreover, we
investigate the effects of the
$\omega-\phi$ mixing which may result in the order of magnitude of the branching ratio $BR(\chi_{c1} \to \omega \phi)$ being $10^{-7}$.
Thus, we are waiting for the accurate measurements of the $BR(\chi_{c1} \to \omega \omega)$, $BR(\chi_{c1} \to \phi \phi)$ and $BR(\chi_{c1}
\to \omega \phi)$, which may be very helpful for
testing the long-distant contribution and the $\omega-\phi$ mixing in $\chi_{c1}\to\phi\phi,\,\omega\omega,\,\omega\phi$ decays.

\end{abstract}

\maketitle

\section{introduction}\label{sec1}

Charm physics is an active field full with chances and challenges
\cite{Li:2008ey}. Decays of charmonia may provide an ideal
laboratory to study perturbative as well as non-perturbative quantum
chromodynamics (QCD). Until now, most of hadronic decays of P-wave
charmonium states $\chi_{cJ}$ ($J=0,1,2$) are not well understood
compared to the $J/\psi$ decays, so that they  cause great interests
of the experimentalists and theorists to further explore the decay
behavior of $\chi_{cJ}$.

In the Hadron 2009 conference, the
BES-III Collaboration announced its observations of $\chi_{cJ}$
($J=0,1,2$) decaying into light vector mesons, where the data of
$\chi_{cJ}$ are taken from $110$ million radiative-decay events of
$\psi(2S)$ collected at the BES-III. Among those decay modes of
$\chi_{cJ}$ decaying into light vector mesons, $\chi_{c1}\to
\phi\phi,\,\omega\omega$ processes were measured for the first time
and the doubly OZI suppressed process $\chi_{cJ}\to \phi\omega$ had
not been measured before the BES-III observation \cite{shencp}.

Generally, the decays of $\chi_{c1}$ into two light vector mesons
are suppressed compared to the corresponding decays of $\chi_{c0}$
and $\chi_{c2}$ due to the helicity selection rule
\cite{Chernyak:1981zz}. Besides, $\chi_{c1}\to \omega\phi$ suffers
from the double-OZI suppression
\cite{Okubo:1963fa,Zweig:1964jf,Zweig:1981pd,Iizuka:1966fk}. Thus,
in the sense, the branching ratios of the channels  $\chi_{1}\to
\phi\phi,\,\omega\omega,\,\phi\omega$ should be small and it would
be difficult to observe them in experiments, especially
$\chi_{c1}\to \phi\omega$. However, the observation of $\chi_{c1}\to
\phi\phi,\,\omega\omega,\, \phi\omega$ seems to be surprising and
compel us to reconsider what mechanism plays the dominant role in
those decays. It would be definitely different from that responsible
for $\chi_{c1}\to \phi\phi,\,\omega\omega,\, \phi\omega$ decays. To
understand the governing mechanism which results in sizable ratios
for $\chi_{c1}\to \phi\phi,\,\omega\omega,\, \phi\omega$, two
questions must be answered: (1) what is the source to alleviate the
helicity selection rule for $\chi_{c1}$; (2) why the double-OZI
suppression is violated for $\chi_{c1}\to \phi\omega$ decay.

The conventional decay mechanism depicting $\chi_{cJ}$ into two
light vector mesons is that $c$ and $\bar{c}$ annihilate into a pair
of gluons, which then transit into quark-antiquark pairs to form the
light vector mesons in the final state. The helicity selection rule
manifests in the $\chi_{cJ}$ decays, and results in the suppression
of $\chi_{c1}$ decaying into two light mesons. In $\chi_{c1}$
decays, the non-perturbative QCD effect plays a crucial role. As an
important non-perturbative effect, the hadronic loop contributions,
which were introduced in Refs. \cite{Lipkin:1986av,Liu:2006dq} and
applied to study charmonium decay
\cite{Liu:2009dr,Zhang:2009kr,Guo:2009wr} and open-charm and hidden
charm decays of charmonium-like states $X,\,Y,\,Z$
\cite{Liu:2006df,Liu:2008yy,Liu:2009iw} extensively, would change
the whole scenario from the conventional decay mechanism of
charmonium.

For $\chi_{c1}$ into two light vector mesons, a quark level
description of the hadronic loop contribution is presented in Fig.
\ref{BW}. Here, red fermion line denotes charm quark, blue and
green lines represent the light quarks. $\chi_{c1}$ first dissolves
into two virtual charmed mesons, then by exchanging an appropriate
hadron (i.e. it possesses appropriate charge, flavor, spin and
isospin) they turn into two on-shell real light hadrons, which can be
caught by detector \cite{Liu:2006dq}. The matrix element of
$\chi_{c1}$ into two light vector mesons via hadronic loop effect
can be described as
\begin{eqnarray}
\mathcal{M}(\chi_{c1}\to VV)=\sum_{i}\langle VV|\mathcal{H}^{(2)}|i\rangle\langle i|\mathcal{H}^{(1)}|\chi_{c1}
\rangle.
\end{eqnarray}
The depiction at the hadron level corresponding to the quark level
diagrams are presented in Fig. \ref{BW}. Here, the suitable
intermediated charmed mesons for the decay of $\chi_{c1}$ into two
light vector mesons should be $D_{(s)}\bar D^{*}_{(s)}+h.c.$, which
interact with $\chi_{c1}$ via S-wave. The exchanged mesons include
pseudoscalar and vector charmed mesons. Thus, by the hadronic loop
mechanism, the transition of $\chi_{c1}$ into two light vector
mesons would not be suppressed by the helicity selection rule.
\begin{widetext}
\begin{center}
\begin{figure}[htb]
\begin{tabular}{cccccccc}
\scalebox{0.5}{\includegraphics{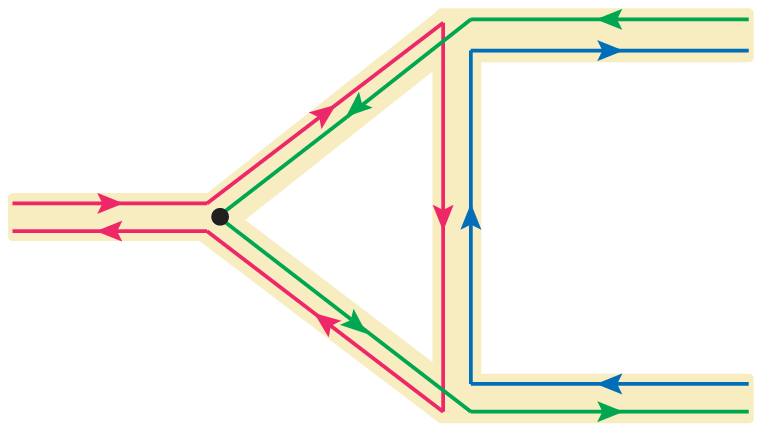}}&\raisebox{7.3ex}{$\Longrightarrow$}&\scalebox{0.7}{\includegraphics{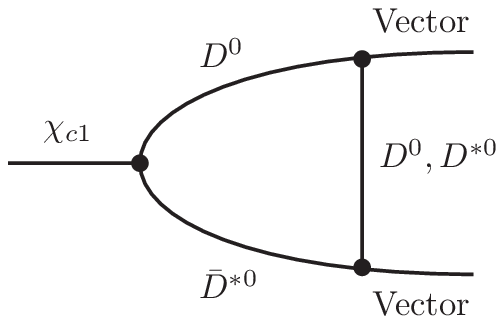}}
&\scalebox{0.7}{\includegraphics{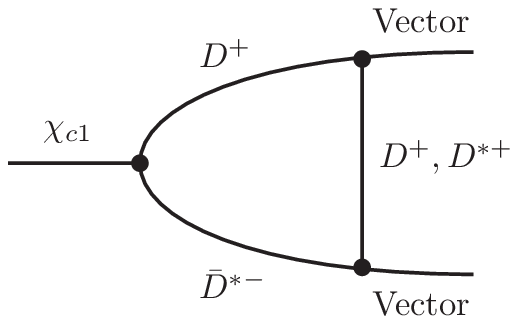}}&\scalebox{0.7}{\includegraphics{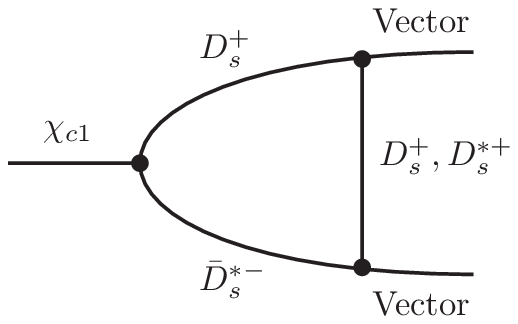}}&\raisebox{7.3ex}{$+ \cdots$}
\end{tabular}
\caption{(Color online). The diagrams of hadronic loop contributions to
$\chi_{c1}\to$ vector$+$vector mesons depicted at the quark level
and hadron level. Here, "Vector" means the light vector meson and
"ellipsis" denotes other diagrams at the hadron level, which can be
obtained by performing a charge conjugation
$D_{(s)}^{(*)}\rightleftharpoons \bar{D}_{(s)}^{(*)}$.\label{BW}}
\end{figure}
\end{center}
\end{widetext}

Another important motivation is how $\chi_{c1}\to \phi\omega$ evades
the double-OZI suppression, assuming $\omega$ and $\phi$ mesons are
ideal mixtures of the flavor $SU(3)$ octet $\omega_8=({u\bar u+d\bar d
-2 s\bar{s}})/{\sqrt{6}}$ and the singlet
$\phi_0=({u\bar{u}+d\bar{d}+s\bar s})/{\sqrt{3}}$. In terms of the
hadronic loop mechanism, such $\chi_{c1}\to \phi\omega$ decay is
fully forbidden. In reality $\omega$ and $\phi$ are not ideal
mixtures of the flavor $SU(3)$ octet and singlet
\cite{Benayoun:1999fv,Kucukarslan:2006wk,Benayoun:1999au,Benayoun:2000ti,Benayoun:2001qz,Benayoun:2007cu},
which would  provide a source which violates the double-OZI
suppression rule for $\chi_{c1}\to \phi\omega$.

In this work, we will combing the hadronic loop effect
with $\omega-\phi$ mixing to study $\chi_{c1}\to \phi\phi,\,\omega\omega,\,\phi\omega$. The paper is organized as follows. After introduction, we present
the formula of hadronic loop contribution to $\chi_{c1}\to
\phi\phi,\,\omega\omega,\,\phi\omega$ with the mixing
schemes for $\omega-\phi$. Then, the numerical results about
$\chi_{c1} \to \omega \omega, \phi\phi, \phi \omega$ are given in Sec. \ref{sec3}. Finally
the paper ends with a discussion and a short summary.

\section{Hadronic loop effect on $\chi_{c1}\to VV$ decays under two different mixing schemes of $\omega-\phi$}\label{sec2}

Firstly, we present the mixing scheme for $\omega-\phi$ used in this
work
\begin{eqnarray}
\left(
  \begin{array}{c}
| \phi^{p} \rangle  \\| \omega^{p}  \rangle\\
  \end{array}
\right)
 =
\left(
  \begin{array}{cc}
    \cos \theta & \sin \theta \\
    -\sin \theta &  \cos \theta
  \end{array}
\right) \left(
  \begin{array}{c}
    | \phi^{I} \rangle \\ | \omega^{I} \rangle \\
  \end{array}
\right),\label{eq0}
\end{eqnarray}
where $|\phi^{p}\rangle$ ($|\omega^{p}\rangle$) and $|\phi^{I}
\rangle$ ($|\omega^{I}\rangle$) are the physical and ideally mixing
states, respectively. The flavor wave functions for the ideally
mixing states $|\omega^I\rangle$ and $|\phi^I\rangle$ are
$\omega^I=(u\bar u+d\bar d)/\sqrt{2}$ and $\phi^I=-s\bar s$. Taking
mixing angle $\theta=0^\circ$ corresponds to the ideal mixing.
According to the analysis in Refs.
\cite{Dolinsky:1991vq,Benayoun:1999fv,Kucukarslan:2006wk}, the
mixing angle $\theta$ should be $(3.4\pm 0.2)^\circ$ in the mixing
scheme in Eq. (\ref{eq0}).

The effective Lagrangians which are responsible for the decay
amplitudes for the diagrams in Fig. \ref{BW}, is listed below
\cite{Cheng:1992xi,Yan:1992gz,Wise:1992hn,Burdman:1992gh,Casalbuoni:1996pg,Colangelo:2003sa}
\begin{eqnarray}
\mathcal{L}_{\chi_{c1}\mathcal{DD}^*}&=&i\,g_{\chi_{c1}\mathcal{DD}^*} {\chi_{c1}}\cdot \mathcal{D}^{*\dag}_i \mathcal{D}^i +h.c.,\\
{\mathcal L}_{H\bar H} =&& i\langle H_b v^\mu {\cal D}_{\mu ba}\bar H_a\rangle+ig\langle
 H_b\gamma_\mu\gamma_5 A_{ba}^\mu\bar H_a\rangle \nonumber \\
 && +i\beta\langle H_b v^\mu(V_\mu-\rho_\mu)_{ba}\bar H_a\rangle\nonumber\\&&+i\lambda\langle
 H_b\sigma^{\mu\nu}F_{\mu\nu}(\rho)_{ba}\bar H_a\rangle,\label{eq1}
\end{eqnarray}
where Eq. (\ref{eq1}) is constructed under the chiral and heavy
quark limits. The superfield $H$ is given by $H={1+v\!\!\!/\over
2}(\mathcal{D}^{*\mu}\gamma_\mu-i\gamma_5 \mathcal{D})$ and $\bar
H=\gamma^0 H^\dag \gamma^0$. $(V_\mu)_{ba}$ and $(A_\mu)_{ba}$
denotes the matrix elements for vector and axial currents,
respectively. The expansion in Eq. (\ref{eq1}), which is related to
the hadronic loop calculation, includes
\begin{eqnarray}
 &&\mathcal{L}_{\mathcal{D}^{(*)}\mathcal{D}^{(*)}\mathcal{V}}\nonumber\\
 &&=-ig_{\mathcal{DD}\mathcal{V}} D_i^\dagger \lrpartial_{\!\mu} D^j(\mathcal{V}^\mu)^i_j\nonumber\\&&
 -2f_{\mathcal{D}^*\mathcal{D}\mathcal{V}} \epsilon_{\mu\nu\alpha\beta}
 (\partial^\mu \mathcal{V}^\nu)^i_j
 (\mathcal{D}_i^\dagger\lrpartial{}^{\!\alpha} \mathcal{D}^{*\beta j}\nonumber\\&&-\mathcal{D}_i^{*\beta\dagger}\lrpartial{}{\!^\alpha} \mathcal{D}^j)
 + ig_{\mathcal{D}^*\mathcal{D}^*\mathcal{V}} \mathcal{D}^{*\nu\dagger}_i \lrpartial_{\!\mu} \mathcal{D}^{*j}_\nu(\mathcal{V}^\mu)^i_j
\nonumber\\
 && +4if_{\mathcal{D}^*\mathcal{D}^*\mathcal{V}} \mathcal{D}^{*\dagger}_{i\mu}(\partial^\mu \mathcal{V}^\nu-\partial^\nu
 \mathcal{V}^\mu)^i_j \mathcal{D}^{*j}_\nu,
\end{eqnarray}
where ${\mathcal{D}^{(*)}}^\dag=(\bar
D^{(*)0},D^{(*)-},D_{s}^{(*)-})$. The coupling constants relevant to
the calculation  include:
$g_{\mathcal{DDV}}=g_{\mathcal{D^*D^*V}}=\beta g_V/\sqrt{2}$,
$f_{\mathcal{D^*DV}}=f_{\mathcal{D^*D^*V}}/m_{\mathcal{D}^*}=\lambda
m_\rho/(\sqrt{2}f_\pi)$ with $\beta=0.9$, $\lambda=0.56$ GeV$^{-1}$
and $f_\pi=132$ MeV
\cite{Cheng:1992xi,Yan:1992gz,Wise:1992hn,Burdman:1992gh}.
$g_{\chi_{c1}\mathcal{DD}^*}=21.4$ GeV for $D$ meson and
$g_{\chi_{c1}\mathcal{DD}^*}=22.6$ GeV for $D_{s}$ meson are determined
in Ref. \cite{Colangelo:2003sa}. Introducing the mixing scheme of
$\omega-\phi$ as shown in Eq. (\ref{eq0}), one defines the $3\times
3$ matrix of the nonet vector mesons $\mathcal{V}$ as
\begin{eqnarray}
\mathcal{V}&=&\left(\begin{array}{ccc}
\frac{\rho^{0}}{\sqrt{2}}+\kappa\omega^p+\zeta \phi^p&\rho^{+}&K^{*+}\\
\rho^{-}&-\frac{\rho^{0}}{\sqrt{2}}+\kappa\omega^p+\zeta \phi^p&
K^{*0}\\
K^{*-} &\bar{K}^{*0}&\delta\omega^p+\sigma \phi^p
\end{array}\right).\nonumber
\end{eqnarray}
If we pre-assume that the mixing
parameters in the matrix are not independent, but related to each
other by a single variable $\theta$, the coefficients
$\kappa,\,\zeta,\,\delta,\,\sigma$ are written as
\begin{eqnarray}
\kappa &=& \frac{1}{\sqrt{2}}\cos \theta ,\quad
\zeta   =  \frac{1}{\sqrt{2}} \sin \theta ,\nonumber\\
\delta &=&  -\sin \theta,\quad\quad\,\,\, \sigma  = \cos \theta.
\end{eqnarray}

Thus, the decay amplitudes of $\chi_{c1}\to
\omega\omega,\,\phi\phi,\,\phi\omega$ due to the hadronic loop
effect are written as
\begin{widetext}
\begin{eqnarray}
&&\mathcal{M}[{\chi_{c1}\to \omega\omega}]\nonumber\\&& =2
\left(\begin{array}{ccc}
\kappa^2&\kappa^2&\delta^2\\
0&0&0\\
0&0&0\end{array}\right)\cdot
\left(\begin{array}{ccc}
\mathcal{M}^{(P)}[D^-,D^{*+},D^{-},\omega,\omega]\\
\mathcal{M}^{(P)}[\bar D^0,{D}^{*0},\bar D^{0},\omega,\omega]\\
\mathcal{M}^{(P)}[D_{s}^-,D_s^{*+},D_s^-,\omega,\omega]\end{array}\right)+2
\left(\begin{array}{ccc}
\kappa^2&\kappa^2&\delta^2\\
0&0&0\\
0&0&0\end{array}\right)\cdot
\left(\begin{array}{ccc}
\mathcal{M}^{(V)}[D^-,D^{*+},D^{*-},\omega,\omega]\\
\mathcal{M}^{(V)}[\bar D^0,{D}^{*0},\bar D^{*0},\omega,\omega]\\
\mathcal{M}^{(V)}[D_{s}^-,D_s^{*+},D_s^{*-},\omega,\omega]\end{array}\right),
\end{eqnarray}
\begin{eqnarray}
&&\mathcal{M}[{\chi_{c1}\to \phi\phi}]\nonumber\\&&=2
\left(\begin{array}{ccc}
0&0&0\\
\zeta^2&\zeta^2&\sigma^2\\
0&0&0\end{array}\right)\cdot
\left(\begin{array}{ccc}
\mathcal{M}^{(P)}[D^-,D^{*+},D^{-},\phi,\phi]\\
\mathcal{M}^{(P)}[\bar D^0,{D}^{*0},\bar D^{0},\phi,\phi]\\
\mathcal{M}^{(P)}[D_{s}^-,D_s^{*+},D_s^-,\phi,\phi]\end{array}\right)+2
\left(\begin{array}{ccc}
0&0&0\\
\zeta^2&\zeta^2&\sigma^2\\
0&0&0\end{array}\right)\cdot
\left(\begin{array}{ccc}
\mathcal{M}^{(V)}[D^-,D^{*+},D^{*-},\phi,\phi]\\
\mathcal{M}^{(V)}[\bar D^0,{D}^{*0},\bar D^{*0},\phi,\phi]\\
\mathcal{M}^{(V)}[D_{s}^-,D_s^{*+},D_s^{*-},\phi,\phi]\end{array}\right),
\end{eqnarray}
\begin{eqnarray}
&&\mathcal{M}[{\chi_{c1}\to \phi\omega}]\nonumber\\&&=
2
\left(\begin{array}{ccc}
0&0&0\\
0&0&0\\
\kappa \zeta& \kappa \zeta & \delta\sigma\end{array}\right)\cdot
\left(\begin{array}{ccc}
\mathcal{M}^{(P)}[D^-,D^{*+},D^{-},\omega,\phi]\\
\mathcal{M}^{(P)}[\bar D^0,{D}^{*0},\bar D^{0},\omega,\phi]\\
\mathcal{M}^{(P)}[D_{s}^-,D_s^{*+},D_s^-,\omega,\phi]\end{array}\right)
+2
\left(\begin{array}{ccc}
0&0&0\\
0&0&0\\
\kappa \zeta& \kappa \zeta & \delta\sigma\end{array}\right)\cdot
\left(\begin{array}{ccc}
\mathcal{M}^{(P)}[D^-,D^{*+},D^{-},\phi,\omega]\\
\mathcal{M}^{(P)}[\bar D^0,{D}^{*0},\bar D^{0},\phi,\omega]\\
\mathcal{M}^{(P)}[D_{s}^-,D_s^{*+},D_s^-,\phi,\omega]\end{array}\right)\nonumber\\&&\quad
+2
\left(\begin{array}{ccc}
0&0&0\\
0&0&0\\
\kappa \zeta& \kappa \zeta & \delta\sigma\end{array}\right)\cdot
\left(\begin{array}{ccc}
\mathcal{M}^{(V)}[D^-,D^{*+},D^{*-},\omega,\phi]\\
\mathcal{M}^{(V)}[\bar D^0,{D}^{*0},\bar D^{*0},\omega,\phi]\\
\mathcal{M}^{(V)}[D_{s}^-,D_s^{*+},D_s^{*-},\omega,\phi]\end{array}\right)+
2
\left(\begin{array}{ccc}
0&0&0\\
0&0&0\\
\kappa \zeta& \kappa \zeta & \delta\sigma\end{array}\right)\cdot
\left(\begin{array}{ccc}
\mathcal{M}^{(V)}[D^-,D^{*+},D^{*-},\phi,\omega]\\
\mathcal{M}^{(V)}[\bar D^0,{D}^{*0},\bar D^{*0},\phi,\omega]\\
\mathcal{M}^{(V)}[D_{s}^-,D_s^{*+},D_s^{*-},\phi,\omega]\end{array}\right),\nonumber\\
\end{eqnarray}
where the factor 2 is from the charge conjugation transformation. In
the above expressions, $\mathcal{M}^{(P)}[\star,\cdots,\star]$ and
$\mathcal{M}^{(V)}[\star,\cdots,\star]$ denote the amplitudes corresponding
to pseudoscalar and vector charmed meson exchanges. The general
expressions of the amplitudes are
\begin{eqnarray}
&&\mathcal{M}^{(P)}[A(p_1),B(p_2),C(q),D(p_3),E(p_4)]\nonumber\\&&\quad=\int\frac{d^4 q}{(2\pi)^4}[ig_{\chi_{c1}\mathcal{DD}^*} \epsilon_\sigma][-ig_{\mathcal{DDV}}(p_1+q)\cdot \epsilon_3][2if_{\mathcal{D^*DV}}\varepsilon_{\mu\nu\alpha\beta}p_4^\mu \epsilon_4^\nu (q^\alpha-p_2^\alpha)]\nonumber\\&&\quad\quad\times
\frac{i}{p_1^2-m^2_\mathcal{D}}\frac{i}{p_2^2-m^2_{\mathcal{D}^*}}\Big(-g^{\sigma\beta}+\frac{p_2^\sigma p_2^\beta}{m^2_{\mathcal{D}^*}}\Big)\frac{i}{q^2-m^2_\mathcal{D}}\,\mathfrak{F}_N^2(q^2,m_{\mathcal{D}}^2),
\end{eqnarray}
\begin{eqnarray}
&&\mathcal{M}^{(V)}[A(p_1),B(p_2),C(q),D(p_3),E(p_4)]\nonumber\\&&\quad=\int\frac{d^4 q}{(2\pi)^4}[ig_{\chi_{c1}\mathcal{DD}^*} \epsilon_\sigma][2if_{\mathcal{D^*DV}}\varepsilon_{\mu\nu\alpha\beta}p_3^\mu \epsilon_3^\nu (p_1^\alpha+q^\alpha)]\nonumber\\&&\qquad\times
\Big[ig_{\mathcal{D^*D^*V}}(q-p_2)\cdot \epsilon_4 g_{\lambda\kappa}-4i f_{\mathcal{D^*D^*V}}(p_{4\kappa}\epsilon_{4\lambda}-p_{4\lambda}\epsilon_{4\kappa})\Big]\nonumber\\&&\qquad\times
\frac{i}{p_1^2-m^2_\mathcal{D}}\frac{i}{p_2^2-m^2_{\mathcal{D}^*}}\Big(-g^{\sigma\kappa}+\frac{p_2^\sigma p_2^\kappa}{m^2_{\mathcal{D}^*}}\Big)\frac{i}{q^2-m^2_{\mathcal{D}^*}}
\Big(-g^{\beta\lambda}+\frac{q^\beta q^\lambda}{m^2_{D^*}}\Big)\mathfrak{F}_N^2(q^2,m_{\mathcal{D}^*}^2)
,
\end{eqnarray}
\end{widetext}
which correspond to pseudoscalar and vector charmed meson
exchanges, respectively. Here, $A$ and $B$ denote the intermediated
charmed mesons. $C$ is the exchanged charmed meson. $D$ and $E$ mean
the light vector mesons in the final states. We adopt the form
factor with the pole form
\begin{eqnarray}
\mathfrak{F}_N(q^2,m^2)=\Bigg(\frac{\Lambda^2-m^2}{\Lambda^2-q^2}\Bigg)^N,\label{pole}
\end{eqnarray}
which depicts the inner structure of the effective vertex of the
exchanged charmed meson and intermediated states. Meanwhile, the
form factor with pole form also plays the role to make the
ultraviolet divergence disappear, in analog to the cut-offs in the
Pauli-Villas renormalization scheme. Here, the cutoff $\Lambda$ can
be parameterized as
$\Lambda=m+\alpha\Lambda_{QCD}$ with $\Lambda_{QCD}=220$ MeV and $m$
is the mass of the exchanged meson \cite{Cheng:2004ru}.

\section{Numerical Results}\label{sec3}

As a free parameter, $\alpha$ is introduced by the cutoff $\Lambda$.
The value is usually dependent on the particular process and taken
to be of the order of unity. The BES-II collaboration reported the
branching ratio of $\chi_{c1}\to K^{*0}(892)\bar{K}^{*0}(892)$ as
$BR[\chi_{c1}\to K^{*0}\bar{K}^{*0}]=(1.67 \pm 0.32 \pm 0.31)\times
10^{-3}$ \cite{Ablikim:2004tv}, which can be applied to determine
$\alpha$ assuming that the hadronic loop effect is dominant in the
process $\chi_{c1}\to K^{*0}\bar{K}^{*0}$. The formula of hadronic
loop contribution to $\chi_{c1}\to K^{*0}\bar{K}^{*0}$ is similar to
that for $\chi_{c1}\to \omega\omega,\,\phi\phi,\,\phi\omega$ decays,
where the exchanged charmed meson matching to intermediated states
$D^{+}\bar D^{*-}+h.c.$ and $D_s^{+}\bar D_s^{*-}+h.c.$ are
$D_s^{(*)+}$ and $D^{(*)}$, respectively. Our study indicates that
the dipole form factor, i.e. taking $N=2$ in Eq. (\ref{pole}), can
well reproduce the branching ratio of  $\chi_{c1}\to
K^{*0}\bar{K}^{*0}$ with  $\alpha= 1.14 \sim 1.28$, which seems to
be reasonable\footnote{In our
calculation, we also tried to fit the measured $B(\chi_{c1}\to
K^{*0}\bar{K}^{*0})$ by adopting the monopole form factor (setting
$N=1$ in Eq. (12)). Although we can also describe the $\chi_{c1}\to
K^{*0}\bar{K}^{*0}$ data, the obtained value of $\alpha$ is far away
from order of unity.  Thus it is more reasonable not to take the
monopole form factor in our calculation. Then, we choose the dipole
form factor instead. In Ref. \cite{Cheng:2004ru}, Cheng et al.
preferred the $N=1$ monopole form factor, whereas, in our work we
adopt $N=2$. The difference between Ref. \cite{Cheng:2004ru} and this
work is due to that the intermediate states of the hadronic loop for
the processes discussed in Ref. \cite{Cheng:2004ru} can be on-shell
while the intermediate states in this work are off-shell.}.

\begin{figure}[h]
\centering
\includegraphics[width=80mm]{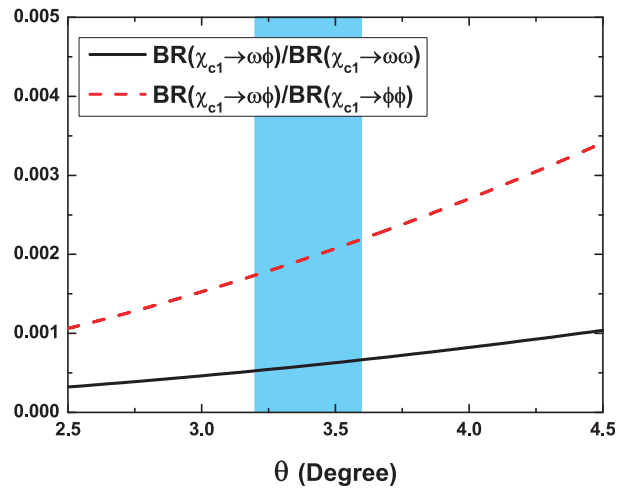}  %
\renewcommand{\figurename}{Fig.}
\caption{(Color online). The $\theta$ dependence of  $BR(\chi_{c1}
\to \omega \phi)/ BR(\chi_{c1} \to \omega \omega)$ and $BR(\chi_{c1}
\to \omega \phi)/ BR(\chi_{c1} \to \phi \phi)$ without $SU(3)_f$
flavor symmetry breaking effect on $\omega-\phi$ mixing. }
\label{nob}%
\end{figure}

\begin{figure}[h]
\centering
\includegraphics[width=80mm]{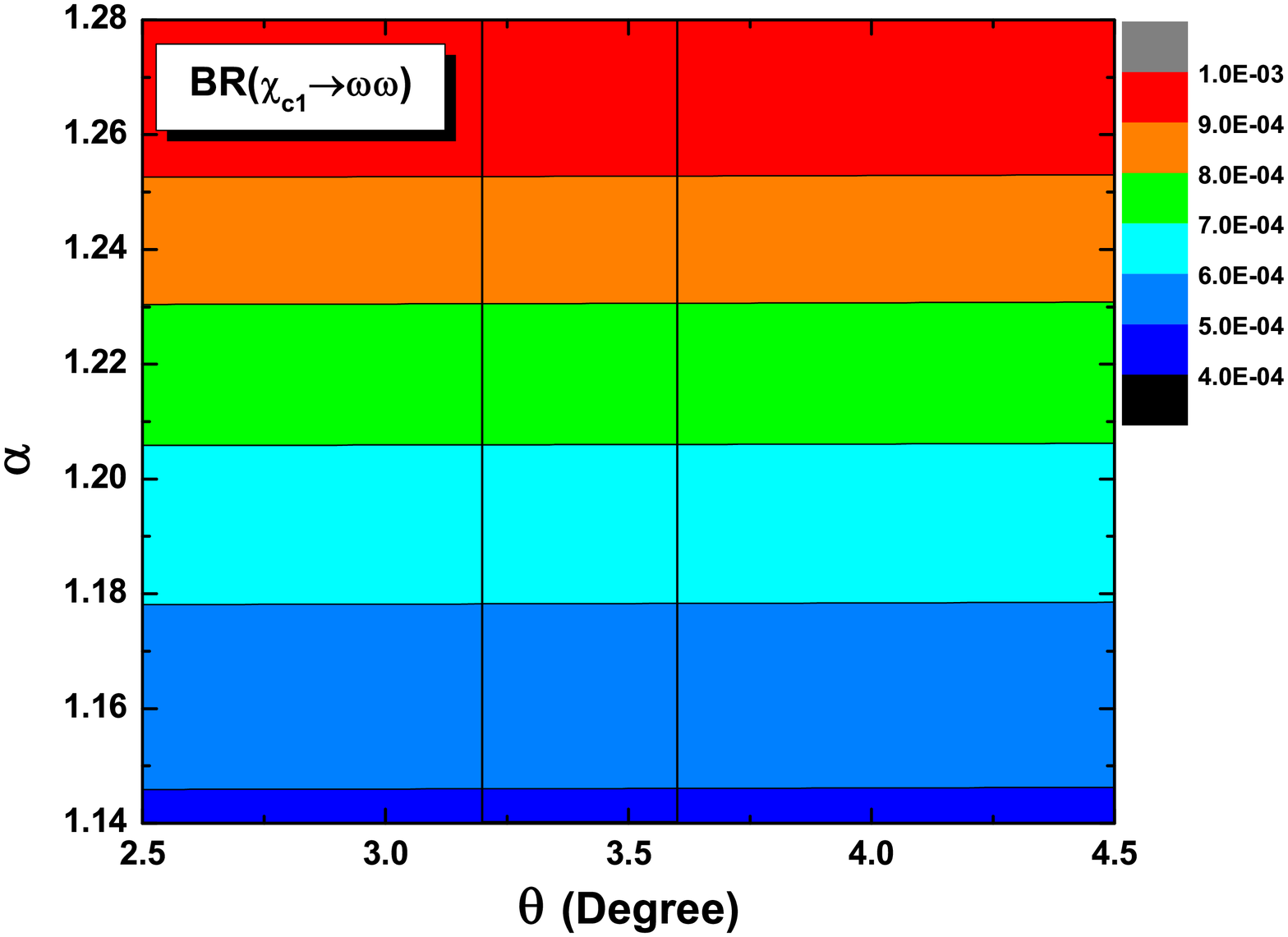}  %
\includegraphics[width=80mm]{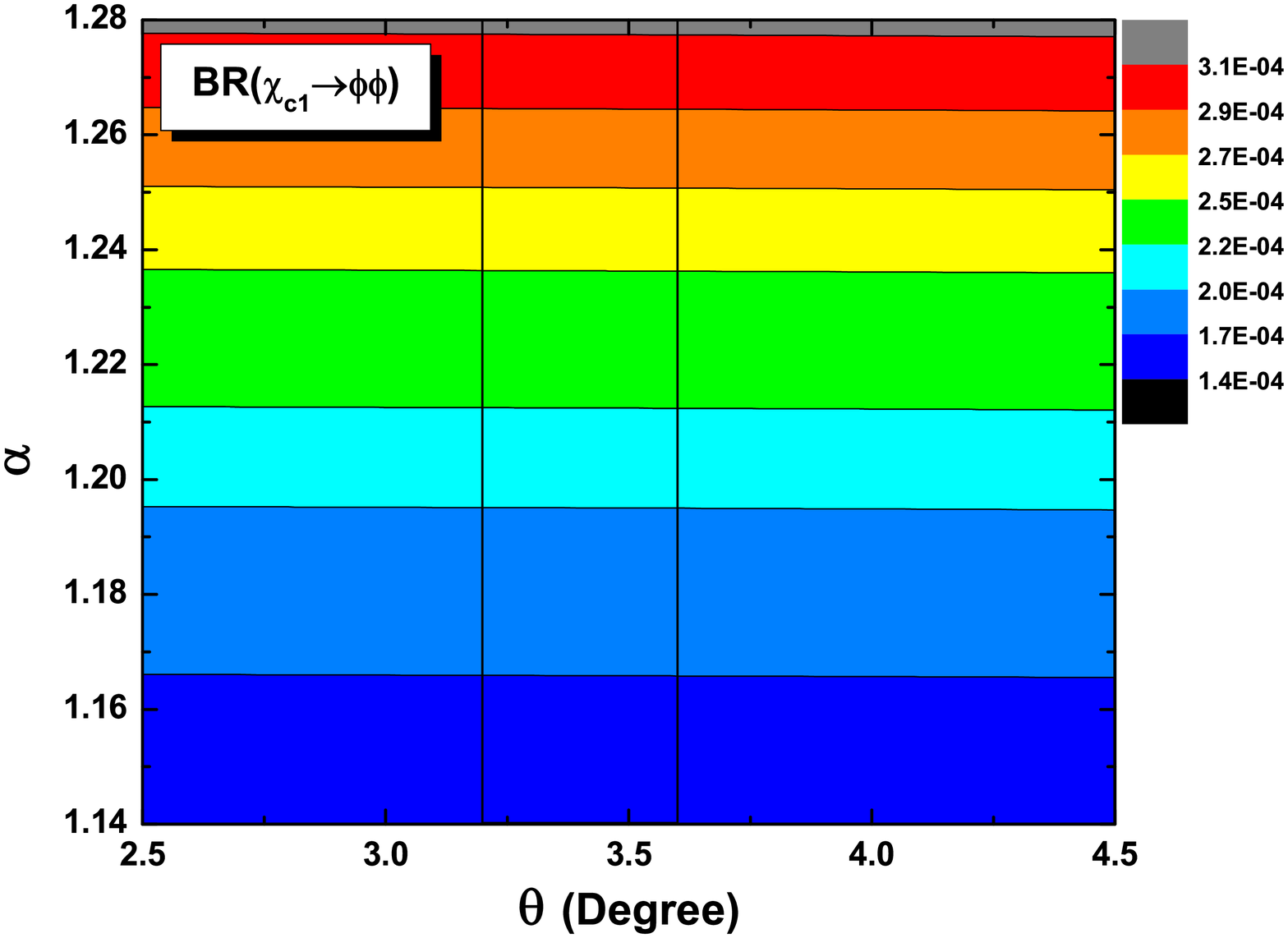}  %
\includegraphics[width=80mm]{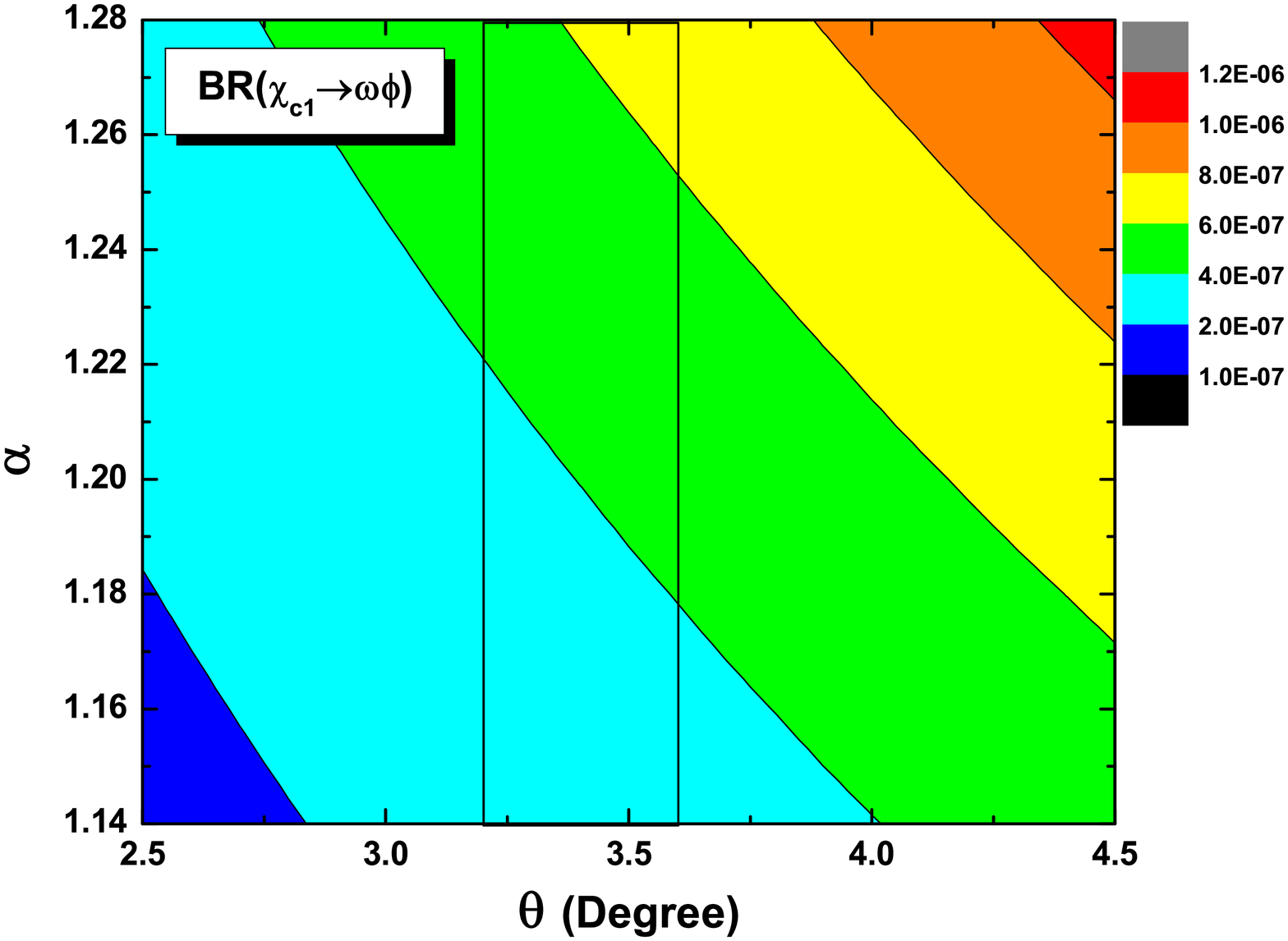}  %
\renewcommand{\figurename}{Fig.}
\caption{(Color online). The contour
plot for the dependence of $BR(\chi_{c1}\rightarrow \omega \omega)$,
$BR(\chi_{c1} \rightarrow \phi \phi)$ and $BR(\chi_{c1}\rightarrow
\omega \phi)$ on $\theta$ and $\alpha$. Here, the range sandwiched
between two vertical solid lines is allowed by the 1-sigma standard deviation
of the mixing angle $\theta$. \label{Contour}}%
\end{figure}

In Fig. \ref{nob} we show the dependence of the ratios of
$BR(\chi_{c1}\to \omega\phi)/BR(\chi_{c1}\to \phi\phi)$ and
$BR(\chi_{c1}\to \omega\phi)/BR(\chi_{c1}\to \omega\omega)$ on
$\theta$ considering the $\omega-\phi$ mixing. In our calculations,
we find these two ratios weakly depend on the parameter $\alpha$. In
the figure a typical value of $\alpha =1.20$ is employed. For the
ideal mixing, i.e. $\theta=0^\circ$, the coefficients $\zeta$ and
$\delta$ are zero, thus, the mesonic loop contribution to the decay
$\chi_{c1} \to \omega \phi$ vanishes. Even though
considering the deviation from the ideal mixing, the ratios of
$BR(\chi_{c1}\to \omega\phi)/BR(\chi_{c1}\to \phi\phi)$ and
$BR(\chi_{c1}\to \omega\phi)/BR(\chi_{c1}\to \omega\omega)$ are
still rather small. In the region $\theta = (3.4\pm 0.2) ^{\circ}$,
the branching ratio of $\chi_{c1} \to \omega \phi$ is three orders
smaller than those of $\chi_{c1} \to \omega \omega$ and $\chi_{c1}
\to \phi \phi$. In Fig.
\ref{Contour}, we present the $\theta$ and $\alpha$ dependence of
the branching ratios of $\chi_{c1}\to \omega\omega$, $\chi_{c1}\to
\phi\phi$ and $\chi_{c1}\to \omega\phi$. The corresponding ranges
of $BR(\chi_{c1}\to \omega\omega)$, $BR(\chi_{c1}\to \phi\phi)$,
$BR(\chi_{c1}\to \omega\omega)$ with $\alpha=1.14\sim 1.28$ and $\theta=(3.4\pm0.2)^\circ$ are listed in the
second column of Table \ref{typical}. In the table, one can notice,
the branching ratio of $\chi_{c1}\to \omega \phi$ is at most of the
order $10^{-7}$.
\par%
\begin{table}[htb]
\begin{center}
\begin{tabular}{c|ccc}\toprule[1pt]
Channel&Branching ratio &Experimental value\\
 $\chi_{c1}\to$&&\\
\midrule[1pt]
$K^{\star 0} \bar{K}^{\star 0}$& $(10.83 \sim 23.25) \times 10^{-4}$ &$(16.7 \pm 3.2 \pm 3.1)\times 10^{-4}$\\
$\omega\omega$& $(4.822 \sim 10.366 )\times 10^{-4}$ &---\\
$\phi\phi$& $(1.465 \sim 3.238) \times 10^{-4} $ &---\\
$\omega\phi$& $(2.542 \sim 6.893) \times 10^{-7}$
&---\\\bottomrule[1pt]
\end{tabular}
\end{center}
\caption{The
ranges of the branching ratios of $\chi_{c1} \rightarrow
\omega\omega,\,\phi\phi,\,\omega\phi$. Here, the experimental data
of $\chi_{c1}\to K^{*0}\bar{K}^{*0}$ provide the central value with
an error tolerance. The range in the $\chi_{c1}\to
K^{*0}\bar{K}^{*0}$ is determined by the error existing in the
experimental data. The experimental error also results in the range
of $\alpha$, i.e. $\alpha=1.14\sim 1.28$. By this determined range
of $\alpha$ and considering the range of mixing angle
($\theta=(3.4\pm0.2)^\circ$), we give the possible range of the
branching ratios of $\chi_{c1} \rightarrow
\omega\omega,\,\phi\phi,\,\omega\phi$.
 \label{typical}}
\end{table}
We need to emphasize that the ratios of $BR(\chi_{c1}\to
\phi\phi)/BR(\chi_{c1}\to \omega\omega)$, $BR(\chi_{c1}\to
\omega\phi)/BR(\chi_{c1}\to \omega\omega)$ and $BR(\chi_{c1}\to
\omega\phi)/BR(\chi_{c1}\to \phi\phi)$ are not sensitive to the
parameter $\alpha$ corresponding to the $\omega-\phi$ mixing.
Furthermore, these two ratios should be independent of the coupling
constants, thus the measurement on their values may provide an ideal
opportunity to test the $\omega-\phi$ mixing.
\par%
\section{Discussion and summary}\label{sec4}

To summarize, in this work, we  study the mesonic loop contributions
and the $\omega-\phi$ mixing effect to the branching ratios of
$\chi_{c1} \to \omega \omega$, $\phi \phi $ and $\omega \phi $. From
the results, one can note that the $\omega-\phi$ mixing plays an
important role in the understanding of the clear signal for
$\chi_{c1} \to \omega \phi$ observed in experiments. Our results
also indicate that accurate measurements on the ratios $BR(\chi_{c1}
\to \omega \phi)/BR(\chi_{c1} \to \omega \omega)$ and $BR(\chi_{c1}
\to \omega \phi)/BR(\chi_{c1} \to \phi \phi)$ are very helpful for
checking mesonic loop contributions and the $\omega-\phi$ mixing
effect.

It is noted from the figures we presented in the text that the
uncertainties in the theoretical computations originate from the
errors of the data, therefore more accurate measurements are
necessary for further studies. Fortunately, a large database on such
rare decay modes will be available at BES-III, which will help
to draw more solid conclusions.

\textit{Notes added}: Very recently, a similar work
\cite{Liu:2009vv} appeared in the arXiv submitted by X.H. Liu and Q.
Zhao when this manuscript was close to completion. In Ref.
\cite{Liu:2009vv}, the authors calculated $\chi_{c1}\to VV$ and
$\chi_{c2}\to VP$ processes by taking hadronic loop effect into
account. Then, the branching ratios of $\chi_{c1}\to
\rho\rho,\,\omega\omega,\,\phi\phi$ are obtained. In our work, we
mainly focused on $\chi_{c1}\to
\omega\omega,\,\phi\phi,\,\omega\phi$ channels, where we also
consider the hadronic loop effect. Our discussion is based on the
recent preliminary results of $\chi_{cJ}\to VV$ presented by the BES
collaboration at the Hadron 2009 conference, especially the first
observation of $\chi_{c1}\to \omega\phi$ \cite{shencp}. Furthermore
we proposed that the $\omega-\phi$ mixing  can be tested via
$\chi_{c1}\to\phi\phi,\,\omega\omega,\,\omega\phi$ decays, which is
different from the idea in Ref. \cite{Liu:2009vv} to some extent.

\section*{Acknowledgements}

X.L. would like to thank Prof. Q. Zhao, who is one of the authors of
Ref. \cite{Liu:2009vv}, for the discussion in the 6th Workshop on
Hadron Physics Theory and the search of New Hadron State. This
project is supported by the National Natural Science Foundation of
China under Grants No. 10705001, No. 10905077 and No. 10775073; the Foundation for
the Author of National Excellent Doctoral Dissertation of P.R. China
(FANEDD) under Contracts No. 200924; the Doctoral Program Foundation of Institutions of
Higher Education of P.R. China under Grant No. 20090211120029; the
Special Grant for the Ph.D program of the Education Ministry of
China; the Program for New Century Excellent Talents in University (NCET) by Ministry of Education of P.R. China under Grant No. NCET-10-0442.

\end{document}